\documentclass[twocolumn,showpacs,preprintnumbers,amsmath,amssymb]{revtex4}

\usepackage{graphicx}
\usepackage{bm}

\begin{document}

\title{Polarized networks, diameter, and synchronizability of networks}

\author{Wei Lin and Xiaowei Zhan}
\affiliation{Department of Mathematics and Statistics, University of
Minnesota Duluth, Duluth, Minnesota 55812, USA}

\date{\today}

\begin{abstract}
Previous research claimed or disclaimed the role of a small diameter
in the synchronization of a network of coupled dynamical systems. We
investigate this connection and show that it is two folds. We first
construct two classes of networks, the polarized networks and the
random networks with a fixed diameter, which exhibit very different
synchronizability. This shows that the diameter itself is
insufficient to determine the synchronizability of networks.
Secondly, we derive analytic estimates on the synchronizability of
networks in terms of the diameter, and find that a larger size of
network admits of a more flexible synchronizability. The analysis is
confirmed by numerical results.
\end{abstract}


\maketitle

\section{Introduction}
Since some critical properties of complex networks were revealed
\cite{pioneerwork}, both theoretical and experimental investigations
of complex networks have received rapidly increasing interests
\cite{review}. After earlier verification of certain characteristics
in various computer-generated and real-world networks \cite{verify},
the focus of research shifted to the dynamics on complex networks
\cite{sync}. The synchronization of a network of dynamical systems
has been extensively analyzed and has found many applications in
secure communication, distance detection, and other engineering
problems \cite{eng}.

It is now well known that the synchronizability of a network can be
characterized by the Laplacian spectrum of its associated graph.
Consider a network of coupled dynamical systems
$\dot{\mathbf{x}}^i=\mathbf{F}(\mathbf{x}^i)+\sigma\sum_{j=1}
^nL_{ij}\mathbf{H}(\mathbf{x}^j)$, where $\mathbf{x}^i$ are the
variables associated with node $i$; $\mathbf{F}$ and $\mathbf{H}$
are evolution and output functions, respectively; $\sigma$ is
control strength; and $L_{ij}$ are entries of the Laplacian matrix,
defined by $L_{ii}=k_i$, the degree of node $i$, $L_{ij}=-1$ if
nodes $i$ and $j$ are connected, and $L_{ij}=0$ otherwise. Order the
eigenvalues of the Laplacian matrix as $0=\lambda_1<\lambda_2
\le\cdots\le\lambda_n$. Then, the larger the two quantities
$\lambda_2$ and $\lambda_2/\lambda_n$ are, the better
synchronizability the network exhibits \cite{sync,sync-eig}.
However, the Laplacian eigenvalues provide little insights into how
the network topology affects the dynamics of synchronization. Thus,
it is important and useful to investigate the connection between the
synchronizability of networks and some intuitive graph invariants.
Since the degree sequence plays a central role in revealing the
scale-free property of networks, its link with synchronizability has
recently been studied \cite{degree}. It was found that the degree
sequence itself is unable to determine the synchronizability of
networks, although it does provide efficient information under
certain homogeneity constraints \cite{homo}. The diameter, which is
defined to be the maximal length of the shortest paths between any
two vertices, has straightforward implication on the structure of a
graph. Some previous experimental studies suggested that a small
diameter might be advantageous to synchronization. This seems quite
reasonable at the first sight, since a small diameter reduces the
distance of signal transmission on the network, but lacks
theoretical verification.

The purpose of this paper is to clarify the connection between the
diameter and the synchronizability of networks. Instead of using
experimental methods directly, we employ both constructions and
analytic estimations. We first explicitly construct a class of
determinate networks, named \textit{polarized networks}, which keeps
the diameter invariant while the network size increases.
Inequalities in spectral graph theory are used to show that
$\lambda_2$ and $\lambda_2/\lambda_n$ approach zero as the network
size increases to infinity, and thus these networks are poorly
synchronizable. Then, we use Erd\H{o}s--R\'{e}nyi random graph model
\cite{erdos59} to obtain a class of random networks with the same
diameter, whose synchronizability proves to be quite good. These
facts are somewhat surprising, since they point out that the
diameter may not be an appropriate quantity to characterize the
synchronizability of networks. To make this observation consistent
with those experiments that suggested the advantage of a small
diameter, we further explore the role that the diameter plays in the
synchronization of networks. Analytic estimates on the
synchronizability of networks in terms of the diameter are derived
and show that the diameter does set lower and upper bounds for
networks of a certain size. By these estimates, we find that the
synchronizability becomes more flexible as the network size
increases. We also present numerical results that support the
theoretical analysis.

\section{Network Constructions}
In this section, we construct two classes of networks with
prescribed diameter but very different synchronizability. To do
this, we first recall some concepts in graph theory \cite{graph}. A
complete graph is a graph with no loops in which each pair of
vertices are joined by one edge. The complete graph of order $n$ is
denoted by $K_n$. Let $A$ and $B$ be two subsets of $V(G)$, the
vertex set of graph $G$, and define the distance between $A$ and $B$
to be the minimal length of all paths between a vertex in $A$ and a
vertex in $B$.

\subsection{Networks with Poor Synchronizability}
Suppose the network size $n$ and the diameter $D$ are given. We
construct the \textit{polarized networks} $P(n, D)$ by joining two
complete graphs $K_{\left\lfloor\frac{n-D+3}{2}\right\rfloor}$ and
$K_{\left\lceil\frac{n-D+3}{2}\right\rceil}$ by a path of length
$D-2$; an example is shown in Fig.\ 1. It is easy to see that
$P(n,D)$ has size $n$ and diameter $D$. Note that the orders of the
two complete graphs differ by at most one, and they are highly
clustering at the ends of a single path, from which the word
``polarized'' comes.

\begin{figure}
\includegraphics[width=.75\columnwidth]{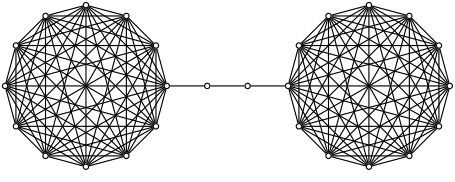}
\caption{An example of polarized networks: $P(26,5)$. Two complete
graphs $K_{12}$ are joined by a path of length 3.}
\end{figure}

To derive an upper bound on the second smallest Laplacian eigenvalue
$\lambda_2$ of $P(n,D)$, we need a result in \cite{alon85}. If $A$
and $B$ are two subsets of $V(G)$ with distance $\rho$, and $F$ is
the set of edges with at least one end vertex not in $A$ or $B$,
then we have the lower bound on the order of $F$:
\begin{equation}\label{lem-F}
|F|\ge\rho^2\lambda_2\frac{|A||B|}{|A|+|B|}.
\end{equation}
Now let $A$ and $B$ be the vertex sets of the two complete graphs in
the construction of $P(n,D)$. Then clearly, we have
$|A|=\left\lfloor\frac{n-D+3}{2}\right\rfloor$,
$|B|=\left\lceil\frac{n-D+3}{2}\right\rceil$, $|F|=D-2$, and
$\rho=D-2$. It follows from (\ref{lem-F}) that
\begin{eqnarray}
\lambda_2&\le&\frac{|F|(|A|+|B|)}{\rho^2|A||B|}=\frac{n-D+3}{(D-2)
\left\lfloor\frac{n-D+3}{2}\right\rfloor\left\lceil\frac{n-D+3}{2}\right\rceil}\nonumber\\
&=&\left\{\begin{array}{ll}\displaystyle\frac{4(n-D+3)}
{(D-2)(n-D+2)(n-D+4)}, & n-D\textrm{ even}\\
\displaystyle\frac{4}{(D-2)(n-D+3)}, & n-D\textrm{ odd}
\end{array}\right.\nonumber\\
&\le&\frac{4}{(D-2)(n-D+3)}.\label{lambda2}
\end{eqnarray}
On the other hand, a lower bound on the largest Laplacian eigenvalue
$\lambda_n$ of a graph was given in \cite{fiedler73}. Let $\Delta$
be the maximal degree of a graph of order $n$. Then
\[
\lambda_n\ge\frac{n}{n-1}\Delta.
\]
Now in $P(n,D)$, $\Delta=\left\lceil\frac{n-D+3}{2}\right\rceil$.
Therefore,
\begin{equation}\label{lambdan}
\lambda_n\ge\frac{n}{n-1}\left\lceil\frac{n-D+3}{2}\right\rceil
\ge\frac{n(n-D+3)}{2(n-1)}.
\end{equation}
Combining (\ref{lambda2}) and (\ref{lambdan}) gives
\begin{equation}\label{ratio}
\frac{\lambda_2}{\lambda_n}\le\frac{8(n-1)}{(D-2)\,n\,(n-D+3)^2}.
\end{equation}
Letting $n\rightarrow\infty$ in (\ref{lambda2}) and (\ref{ratio})
yields that both $\lambda_2$ and $\lambda_2/\lambda_n$ approach
zero, which implies that the synchronizability of polarized networks
can be arbitrarily poor as the network size increases. What is more,
we will show in the next section that the polarized networks are
nearly the optimal class of networks with poorest synchronizability
among all the networks with the same network size and diameter.

\subsection{Networks with Good Synchronizability}
Instead of explicitly constructing a determinate network, we
construct a class of random networks based on Erd\H{o}s--R\'{e}nyi
random graph model. By this construction, we show that although
there exist some extremal cases like the polarized networks, a
general class of networks has fairly good synchronizability.

A random graph $G(n,p)$ consists of $n$ vertices and the edges
chosen independently with probability $p$. For the purpose of our
construction, we need to fix the diameter of the graphs for all
network size $n$. However, the diameter of $G(n,p)$ may vary when
$n$ increases, depending on how we choose $p$ as a function of $n$.
Thus, we need a result concerning the diameter of $G(n,p)$
\cite{bollobas01}. Suppose the functions $D=D(n)\ge3$ and
$0<p=p(n)<1$ satisfy that $\log n/D-3\log\log n\rightarrow\infty$,
$p^Dn^{D-1}-2\log n\rightarrow \infty$, and $p^{D-1}n^{D-2}-2\log
n\rightarrow-\infty$. Then $G(n,p)$ almost surely, i.e.\ with
probability tending to one as $n$ tends to infinity, has diameter
$D$. Here, let $D\ge3$ be a constant and
\begin{equation}\label{pn}
p=\frac{(\alpha\,n\log n)^{1/(D-\varepsilon)}}{n}
\end{equation}
where $0<\varepsilon<1$ and $\alpha$ can be any positive number. It
is easy to verify that all the conditions mentioned above are
satisfied, which yields that a random graph $G(n,p)$ with $p$
defined in (\ref{pn}) almost surely has diameter $D$.

The random graphs have a very narrow degree distribution. In fact,
we have the proposition \cite{krivelevich05}, saying that if
$p=p(n)$ satisfies that $pn/\log n\rightarrow\infty$ and
$(1-p)\,n\log n\rightarrow\infty$, then almost surely all the
degrees of $G(n,p)$ are equal to $(1+o(1))np$. It is clear that the
assumptions are satisfied under (\ref{pn}). Hence, almost surely all
the degrees of $G(n,p)$ with (\ref{pn}) holding are near $np$.

Now we are concerned with the Laplacian eigenvalues associated with
$G(n,p)$. The Laplacian $L$ and the adjacency matrix $A$ of a graph
are linked by the relation $L=K-A$, where $K$ is the diagonal matrix
with diagonal entries being the vertex degrees. Due to the narrow
distribution of degrees in $G(n,p)$, we have that $L\simeq npI-A$,
and so
\begin{equation}\label{lam-lambar}
\lambda_i\simeq np-\bar{\lambda}_i,\quad i=\overline{1,n},
\end{equation}
where $\bar{\lambda}_i$ denote the eigenvalues of $A$. Therefore,
what we need to know is the distribution of eigenvalues of the
adjacency matrix $A$. We know that this distribution follows a
semicircle law \cite{wigner}, and a further result concerning all
the eigenvalues $\bar{\lambda}_i$ except $\bar{\lambda}_1$
\cite{furedi81} says that
\begin{equation}\label{lambarbound}
\max_{2\le i\le n}\left|\bar{\lambda}_i(G(n,p))\right|=O(\sqrt{np}).
\end{equation}
And also notice that $\bar{\lambda}_1$ stands far away from the bulk
of the spectrum, at $np$. Combining (\ref{lam-lambar}) and
(\ref{lambarbound}) yields
\[
\lambda_i\simeq np+O(\sqrt{np}),\quad i=\overline{2,n}.
\]
With (\ref{pn}), we obtain the approximations for the two quantities
characterizing the synchronizability
\begin{equation}\label{lambda2r}
\lambda_2\simeq np=(\alpha\,n\log n)^{1/(D-\varepsilon)},
\end{equation}
\begin{equation}\label{ratior}
\frac{\lambda_2}{\lambda_n}\simeq\frac{np-O(\sqrt{np})}
{np+O(\sqrt{np})},
\end{equation}
which implies that $\lambda_2\rightarrow\infty$ and
$\lambda_2/\lambda_n\rightarrow1$ as $n\rightarrow\infty$. By this
we have shown that the class of random networks we constructed keeps
the diameter invariant but exhibits rather good synchronizability.

\section{Analytic Estimations}
By the constructions given in the previous section, we see that the
synchronizability of networks with the same diameter may differ
significantly. They also suggest that the diameter should not be an
appropriate quantity to characterize the synchronizability of
networks, although this is somewhat conflictive with intuition.
Naturally, one could ask whether or not there exists some relation
between the diameter and the synchronizability. The answer is yes;
we will now confirm this relation by deriving some estimates on the
quantities $\lambda_2$ and $\lambda_2/\lambda_n$. These estimates
are based on a series of inequalities in spectral graph theory.

A lower bound on $\lambda_2$ in terms of the network size $n$ and
the diameter $D$ \cite{mohar91} pointed out that
\begin{equation}\label{lamda2low}
\lambda_2\ge\frac{4}{nD}.
\end{equation}
Chung \cite{chung94} estimated the diameter $D$ by giving the upper
bound
\[
D\le\left\lfloor\frac{\cosh^{-1}(n-1)}{\cosh^{-1}\left(
\frac{\lambda_n+\lambda_2}{\lambda_n-\lambda_2}\right)}\right\rfloor+1,
\]
which implies that
\[
D-1\le\frac{\cosh^{-1}(n-1)}{\cosh^{-1}\left(\frac{1+\lambda_2/\lambda_n}
{1-\lambda_2/\lambda_n}\right)}.
\]
We can solve this inequality and obtain an upper bound on
$\lambda_2/\lambda_n$
\begin{equation}\label{ratioup}
\frac{\lambda_2}{\lambda_n}\le\frac{\cosh\left(\frac{\cosh^{-1}(n-1)}
{D-1}\right)-1}{\cosh\left(\frac{\cosh^{-1}(n-1)}{D-1}\right)+1}.
\end{equation}
Now note that a simple upper bound on $\lambda_n$ was given
\cite{kelmans67} by
\begin{equation}\label{lambdanup}
\lambda_n\le n
\end{equation}
with equality holding if and only if the complement of the graph is
disconnected. Combining (\ref{lamda2low}), (\ref{ratioup}), and
(\ref{lambdanup}) gives the estimates as follows.

\textit{Proposition: Given the network size $n$ and the diameter
$D$, the following estimates on $\lambda_2$ and
$\lambda_2/\lambda_n$ hold:}
\begin{equation}\label{prop1}
\frac{4}{nD}\le\lambda_2\le
n\,\frac{\cosh\left(\frac{\cosh^{-1}(n-1)}{D-1}\right)
-1}{\cosh\left(\frac{\cosh^{-1}(n-1)}{D-1}\right)+1},
\end{equation}

\begin{equation}\label{prop2}
\frac{4}{n^2D}\le\frac{\lambda_2}{\lambda_n}\le\frac{\cosh\left(
\frac{\cosh^{-1}(n-1)}{D-1}\right)
-1}{\cosh\left(\frac{\cosh^{-1}(n-1)}{D-1}\right)+1}.
\end{equation}

\textit{Proof:} The upper bound on $\lambda_2$ follows from
(\ref{ratioup}) and (\ref{lambdanup}), and the lower bound on
$\lambda_2/\lambda_n$ follows from (\ref{lamda2low}) and
(\ref{lambdanup}).\hfill$\Box$

To gain more insights into the connection between the diameter and
the synchronizability, we find an asymptotic order of the upper
bound on $\lambda_2/\lambda_n$. It is easy to calculate that
\[
\frac{\cosh\left(\frac{\cosh^{-1}(n-1)}{D-1}\right)
-1}{\cosh\left(\frac{\cosh^{-1}(n-1)}{D-1}\right)+1}\simeq
\frac{(2n)^{1/(D-1)}-2}{(2n)^{1/(D-1)}+2}.
\]
With the estimates and asymptotic order, it is easily seen that a
larger diameter $D$ implies both smaller lower bounds and smaller
upper bounds on $\lambda_2$ and $\lambda_2/\lambda_n$. This
observation coincides with the intuition that a larger diameter is
likely to be harmful to synchronization. From this point of view,
the assertion that a small diameter benefits synchronization could
be true, although it is not true for individual networks.

Another observation worthwhile to point out is that a larger network
size $n$ implies smaller lower bounds and larger upper bounds on
both $\lambda_2$ and $\lambda_2/\lambda_n$. In other words, the
synchronizability of networks becomes more flexible as the network
size increases. This fact suggests that the indicator role of the
diameter in synchronization will substantially weaken in large-size
networks.

Let us return to the polarized networks for which the estimates on
$\lambda_2$ and $\lambda_2/\lambda_n$ were given in (\ref{lambda2})
and (\ref{ratio}), respectively. The asymptotic orders of the upper
bounds are $4/(nD)$ and $8/(n^2D)$, respectively, for large $n$ and
$D$, and $n\gg D$. Thus, by the Proposition, we see that the
polarized networks are almost the optimal class of networks with
poorest synchronizability.

\section{Numerical Results}
In this section, we carry out numerical experiments to explore the
connection between the diameter and the synchronizability for the
two classes of networks we constructed. Note that the polarized
networks are determinate while the random networks are, of course,
not. So the former case is much simpler, while some average methods
should be used to eliminate random perturbations in the latter case.

In the polarized networks, all we need to determine before
generating the network topology are the network size $n$ and the
diameter $D$. To investigate the influence of network size $n$ and
diameter $D$ on the synchronizability of polarized networks, we
assigned a set of values to the two parameters, and plotted the
trends of the two quantities that characterize the synchronizability
in Figs.\ 2 and 3. According to the lower bound estimates in
(\ref{prop1}) and (\ref{prop2}), combined with the upper bound
estimates in (\ref{lambda2}) and (\ref{ratio}), the relationships
between the pairs of parameters we have shown in Figs.\ 2 and 3 are
expected to be all linear. The numerical results are quite
consistent with and support this observation. They confirm that the
synchronizability of polarized networks is very poor, and decreases
toward zero at least linearly as the networks size and diameter
increase.

\begin{figure}
\includegraphics[width=\columnwidth]{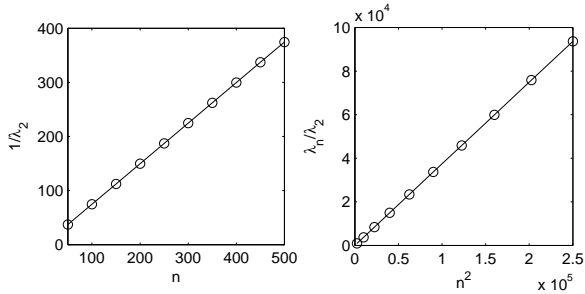}
\caption{The relationships between the network size $n$ and the two
quantities that characterize the synchronizability of polarized
networks $P(n,D)$. The diameter $D$ is set to 5 all the time.}
\end{figure}

\begin{figure}
\includegraphics[width=\columnwidth]{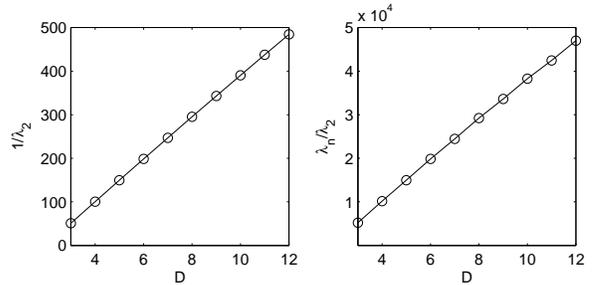}
\caption{The relationships between the diameter $D$ and the two
quantities that characterize the synchronizability of polarized
networks $P(n,D)$. The network size $n$ is set to 200 all the time.}
\end{figure}

Now we turn to the random networks $G(n,p)$ with $p$ defined in
(\ref{pn}). It is worth noticing that although any $0<\varepsilon<1$
and $\alpha>0$ ensure the convergence of diameter to the specified
value, a careful selection of their values will accelerate the
convergence so that networks with a particular diameter may be
obtained for relatively small network sizes. In our experiments, we
always set $\varepsilon=0.5$, but choose different $\alpha$ values
for different diameters $D$. The results of experiments for $D=4$
and $\alpha=5$ are illustrated in Fig.\ 4. In the experiments, ten
random networks were generated for each network size $n$, and the
averages of $\lambda_2$ and $\lambda_2/\lambda_n$ were computed
respectively. The diameter converged fast to and stayed stably at
the specified value. This is shown in the bottom of Fig.\ 4, where
$N$ denotes the number of networks achieving diameter $D$ out of the
ten for each network size $n$. Meanwhile, the two quantities
$\lambda_2$ and $\lambda_2/\lambda_n$ related to the
synchronizability went up slowly. These trends are in agreement with
the analytic estimations. However, it is difficult to obtain the
accurate growing speeds of the two quantities by numerical
experiments, since the required computations increase very rapidly
along with the network size.

\begin{figure}
\includegraphics[width=\columnwidth]{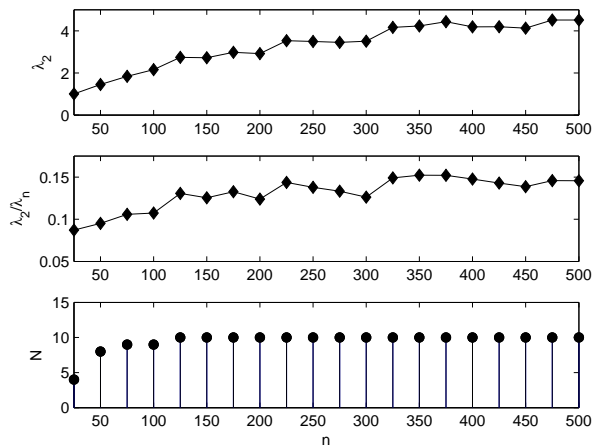}
\caption{The performance of random networks with $p$ defined in
(\ref{pn}). For the set of fixed parameters $D=4$,
$\varepsilon=0.5$, $\alpha=5$, ten random networks were generated
for each network size $n$. The averages of the two quantities
related to the synchronizability are shown in the top two graphs,
while $N$, the number of networks achieving diameter $D$ out of the
ten for each network size $n$ is shown in the bottom.}
\end{figure}

\begin{figure}
\includegraphics[width=\columnwidth]{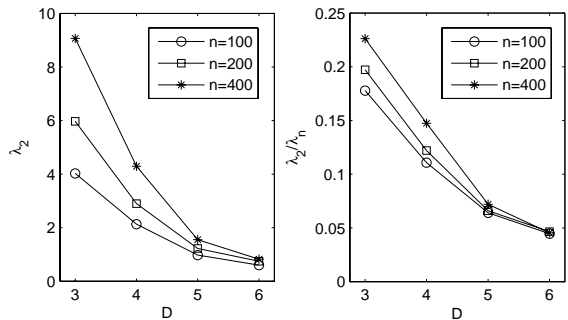}
\caption{The average performance of the class of random networks we
constructed, under different diameters and network sizes. Here
$\varepsilon=0.5$, and $\alpha$ is set appropriately to ensure fast
convergence to the specified diameter. For each set of parameters,
50 random networks were generated and averages of the two quantities
related to the synchronizability were computed respectively.}
\end{figure}

More accurate numerical results are shown in Fig.\ 5, where 50
random networks were generated for each set of parameters and
averages of $\lambda_2$ and $\lambda_2/\lambda_n$ were computed
respectively, so that the effects of random perturbations could be
reduced remarkably. First, it is clear that the increasing diameter
has pulled down the synchronizability of networks. This is
consistent with the intuitive observation that a large diameter
would impede communications between nodes and synchronization in a
certain class of random networks, although the observation is
absolutely not true for determinate networks, for instance, a
polarized network and a typical sample in the random networks with
the same diameter, which have been shown to perform distinctly in
synchronization.

Another observation which can be seen from Fig.\ 5 is that the class
of random networks has an obvious trend in the improvement of
synchronizability as the network size increases, which turns out to
be the same as what has been confirmed by Fig.\ 4. However, it is
worth noticing that this trend becomes quite faint for a large
diameter. This can be seen from the asymptotic orders in
(\ref{lambda2r}) and (\ref{ratior}). Restricted by the complexity of
computation, we were only able to conduct the experiments in a
relatively small range of network size $n$, but the increase of
network size was pulled down considerably by a large diameter $D$,
according to (\ref{lambda2r}) and (\ref{ratior}).

\section{Conclusions}
We investigated the connection between the diameter and the
synchronizability of networks. By constructing determinate polarized
networks and a class of random networks, we found that the networks
with the same diameter may have very different synchronizability.
Thus, the diameter itself is inappropriate to be a quantity that
characterizes the synchronizability of networks. We also derived
analytic estimates on the range of synchronizability of networks
with specified diameter and network size, which revealed that a
larger size of network admits of a more flexible synchronizability.

An application of the network constructions is the design of network
topology in communications, multi-agent systems, and other
engineering problems. Since the polarized networks are almost the
optimal class of networks with poorest synchronizability, they are
especially suitable to be used in the design of easily
desynchronizable networks. Moreover, the nature of explicit
constructions makes it quite easy to apply. The random networks with
fixed diameter are also useful in synchronization-related
applications that have special requirements on the diameter of
networks. The fast convergence to the specified diameter enables us
to generate desired networks of relatively small sizes.

A noteworthy endeavor in the literature is to explicitly construct
networks with best synchronizability. Recently, methods in numerical
optimization such as simulated annealing have been utilized to
obtain nearly optimal network topology \cite{donetti05}. It was
suggested that the optimal network topology should possess several
important properties, including homogeneous degree, betweenness, and
distance distributions, large girths, small diameters, no community
structure, etc. However, the procedure to generate such a network
topology remains an open question. As pointed out in this paper and
previous research, individual properties such as small diameters,
homogeneous degree distributions, etc., are insufficient to optimize
the synchronizability of networks. A successful constructing
algorithm should instead take into account a full combination of
these favorable properties.

\section*{ACKNOWLEDGMENTS}
The authors thank Dalibor Fron\v{c}ek and Zhuangyi Liu for their
encouragement and helpful discussions.


\begin{thebibliography}{99}
\bibitem{pioneerwork} D.~J. Watts and S.~H. Strogatz, Nature
(London) \textbf{393}, 440 (1998); A.-L. Barab{\'a}si and R. Albert,
Science, \textbf{286}, 509 (1999).

\bibitem{review} S.~H. Strogatz, Nature (London) \textbf{410}, 268
(2001); R. Albert and A.-L. Barab{\'a}si, Rev.\ Mod.\ Phys.
\textbf{74}, 47 (2002); M.~E.~J. Newman, SIAM Rev. \textbf{45}, 167
(2003); S. Boccaletti \textit{et al.}, Phys.\ Rep. \textbf{424}, 175
(2006).

\bibitem{verify} H. Jeong \textit{et al.}, Nature (London)
\textbf{407}, 651 (2000); L.~A.~N. Amaral \textit{et al.}, Proc.\
Natl.\ Acad.\ Sci.\ U.S.A. \textbf{97}, 11149 (2000); R.
Pastor-Satorras, A. V{\'a}zquez, and A. Vespignani, Phys.\ Rev.\
Lett. \textbf{87}, 258701 (2001).

\bibitem{sync} X.~F. Wang and G. Chen, Int.\ J. Bifurcation Chaos
Appl.\ Sci.\ Eng. \textbf{12}, 187 (2002); M. Barahona and L.~M.
Pecora, Phys.\ Rev.\ Lett. \textbf{89}, 054101 (2002).

\bibitem{eng} X. Li and G. Chen, IEEE Trans.\ Circuits Syst.\ I:
Fundam.\ Theory Appl. \textbf{50} 1381 (2003); W. Lu and T. Chen,
Physica D \textbf{213}, 214 (2006).

\bibitem{sync-eig} C.~W. Wu and L.~O. Chua, IEEE Trans.\ Circuits
Syst.\ I: Fundam.\ Theory Appl. \textbf{42}, 430 (1995); J. Jost and
M.~P. Joy, Phys.\ Rev.\ E \textbf{65}, 016201 (2002); F.~M. Atay, J.
Jost, and A. Wende, Phys.\ Rev.\ Lett. \textbf{92}, 144101 (2004);
W. Lu and T. Chen, Physica D \textbf{198}, 148 (2004).

\bibitem{degree} F.~M. Atay, T. B{\i}y{\i}ko\u{g}lu, and J. Jost,
IEEE Trans.\ Circuits Syst.\ I: Regular Papers \textbf{53}, 92
(2006); C.~W. Wu, Phys.\ Lett.\ A \textbf{346}, 281 (2005).

\bibitem{homo} T. Nishikawa, A.~E. Motter, Y.-C. Lai, and F.~C.
Hoppensteadt, Phys.\ Rev.\ Lett. \textbf{91}, 014101 (2003); F.
Chung, L. Lu, and V. Vu, Proc.\ Natl.\ Acad.\ Sci.\ U.S.A.
\textbf{100}, 6313 (2003).

\bibitem{erdos59} P. Erd\H{o}s and A. R{\'e}nyi, Publ.\ Math.\
(Debrecen) \textbf{6}, 290 (1959).

\bibitem{graph} R. Diestel, \textit{Graph Theory} (Springer,
Heidelberg, 2005), http://
www.math.uni-hamburg.de/home/diestel/books/graph.theory.

\bibitem{alon85} N. Alon and V.~D. Milman, J. Combin.\ Theory, Ser.\
B \textbf{38}, 73 (1985).

\bibitem{fiedler73} M. Fiedler, Czech.\ Math.\ J. \textbf{23}, 298
(1973).

\bibitem{bollobas01} B. Bollob{\'a}s, \textit{Random Graphs}
(Cambridge University Press, Cambridge, 2001), p.~263.

\bibitem{krivelevich05} M. Krivelevich and B. Sudakov, e-print
math.CO/0503745.

\bibitem{wigner} E.~P. Wigner, Ann.\ Math. \textbf{62}, 548 (1955);
\textbf{67}, 325 (1958).

\bibitem{furedi81} Z. F{\"u}redi and J. Koml{\'o}s, Combinatorica
\textbf{1}, 233 (1981).

\bibitem{mohar91} B. Mohar, Graphs Combin. \textbf{7}, 53 (1991).

\bibitem{chung94} F.~R.~K. Chung, V. Faber, and T.~A. Manteuffel,
SIAM J. Discrete Math. \textbf{7}, 443 (1994).

\bibitem{kelmans67} A.~K. Kel'mans, in \textit{Cybernetics---in the
Service of Communism}, Vol.~4 (Izdat.\ ``{\`E}nergija'', Moscow,
1967), pp.~27--41 (Russian).

\bibitem{donetti05} L. Donetti, P.~I. Hurtado, and M.~A. Mu{\~n}oz,
Phys.\ Rev.\ Lett. \textbf{95}, 188701 (2005).
\end{thebibliography}
\end{document}